\documentclass[10pt,conference,compsocconf]{IEEEtran}

\usepackage{cite}
\usepackage{amsmath,amssymb,amsfonts}
\usepackage{algorithmic}
\usepackage{graphicx}
\usepackage{textcomp}
\usepackage{xcolor}
\usepackage{array}
\usepackage{algorithm}
\usepackage{hyperref}
\usepackage{flafter}
\usepackage{placeins}

\usepackage{booktabs}
\newcommand{\Description}[1]{} 
\usepackage{subcaption}
\usepackage{xurl}
\usepackage{url}
\begin{document}
\bstctlcite{BSTcontrol}

\title{DySCo: Dynamic Sharding for Collaborative Edge–Cloud LLM Inference with Depth-Synchronized Batching}

\author{
\IEEEauthorblockN{Jingpo Xu}
\IEEEauthorblockA{
University of Amsterdam\\
Amsterdam, Netherlands\\
j.xu3@uva.nl
}
\and
\IEEEauthorblockN{Paul Joe Maliakel}
\IEEEauthorblockA{
TU Wien\\
Vienna, Austria\\
paul.maliakel@tuwien.ac.at
}
\and
\IEEEauthorblockN{Ivona Brandic}
\IEEEauthorblockA{
TU Wien\\
Vienna, Austria\\
ivona.brandic@tuwien.ac.at
}
\and
\IEEEauthorblockN{Shashikant Ilager}
\IEEEauthorblockA{
University of Amsterdam\\
Amsterdam, Netherlands\\
s.s.ilager@uva.nl
}
}

\maketitle

\begin{abstract}
Pervasive intelligent applications are increasingly deployed on mobile and Internet of Things (IoT) edge devices. Consequently, Large Language Models (LLMs) are increasingly used to support these applications. Yet, due to their high resource demands, LLMs are mostly deployed in the cloud. Layer-wise edge--cloud inference lets resource-constrained edge devices contribute computation to LLMs they cannot host in full. However, heterogeneous split points introduce two coupled inefficiencies. \textit{First}, edge execution and communication create idle gaps between cloud invocations. \textit{Second}, requests arriving at different model depths cannot be conventionally batched. We present DySCo, a collaborative runtime that keeps KV caches local and introduces \emph{dyForward}, a model-aware layer-range executor that runs configurable contiguous layer ranges from resident model shards without reloading weights. For multi-edge serving settings, we introduce depth-synchronized batching (DSB), which advances heterogeneous requests to the deepest cut and batches their common suffix. 
Experiments across heterogeneous devices, two model families, and local and wide-area links show that idle gaps increase the latency of subsequent GPU forward calls even when waiting time is excluded, adding up to 25\,ms of additional cloud-side suffix latency per decoding step in our measurements. At an average concurrency of eight, DSB improves throughput by $275\%$ over FIFO, $48\%$ over exact-match batching, and $79\%$ over round-robin interleaving while reducing mean per-session latency. Together, these results show that requests with different edge--cloud splits can reuse resident cloud weights and share batched suffix computation.
The artifact repository for this work is publicly available at
\url{https://github.com/Large-scale-Sustainable-Computing-LSC/dysco-artifact}.
\end{abstract}


\section{Introduction}

Large language models (LLMs) are increasingly used in interactive and
personal applications, yet their inference remains difficult to place close to users. Cloud deployment provides sufficient memory, accelerator capacity, and mature serving optimizations, and therefore remains the natural choice for hosting LLMs. However, it requires the entire inference to be performed remotely, leaving the available resources on edge devices outside the inference path. In contrast, fully local deployment relies exclusively on edge resources, whose limited memory, compute, and energy budgets restrict the model sizes they can host, even with model compression and edge-oriented architectures~\cite{A_Review_rw, liu2024mobilellmoptimizingsubbillionparameter_1_24}. The resulting gap motivates an intermediate design point in which an edge device contributes its available resources by executing part of the model, while the cloud completes the remaining computation.

Prior systems route requests between smaller and larger models~\cite{ong2025routellm}, or separate prefill and decode across servers~\cite{splitwise2024,distserve2024}. These approaches generally assign complete requests or execution phases to devices capable of hosting the required model. We instead consider resource-constrained edge devices that cannot host the full model, and partition each inference request between the edge and cloud at a configurable layer boundary. In this layer-wise collaborative setting, the edge executes a model prefix and passes its hidden state to a cloud-resident suffix. This approach can pool memory and computation without changing model weights, enabling distributed model execution. Prior systems have established the feasibility of layer-wise cloud--edge partitioning and distributed LLM shards~\cite{Neurosurgeon_dis,Petals-dis,edgeshard_rw,splitllm2024}.

Consider a fleet of edge devices serving the same model. A device with available compute can execute a longer model prefix, while a busy or more resource--constrained device can execute a shorter one. For a new session, the edge selects how many layers to execute from its available prefix shard; the cloud executes the corresponding suffix from a shared model instance. This flexibility avoids moving model weights during inference, but causes requests to reach the cloud at different times and model depths. The cloud must therefore handle both the gaps between ready requests and the loss of conventional batching across different cuts.

\begin{figure}[t]
  \centering
  \includegraphics[width=\linewidth]{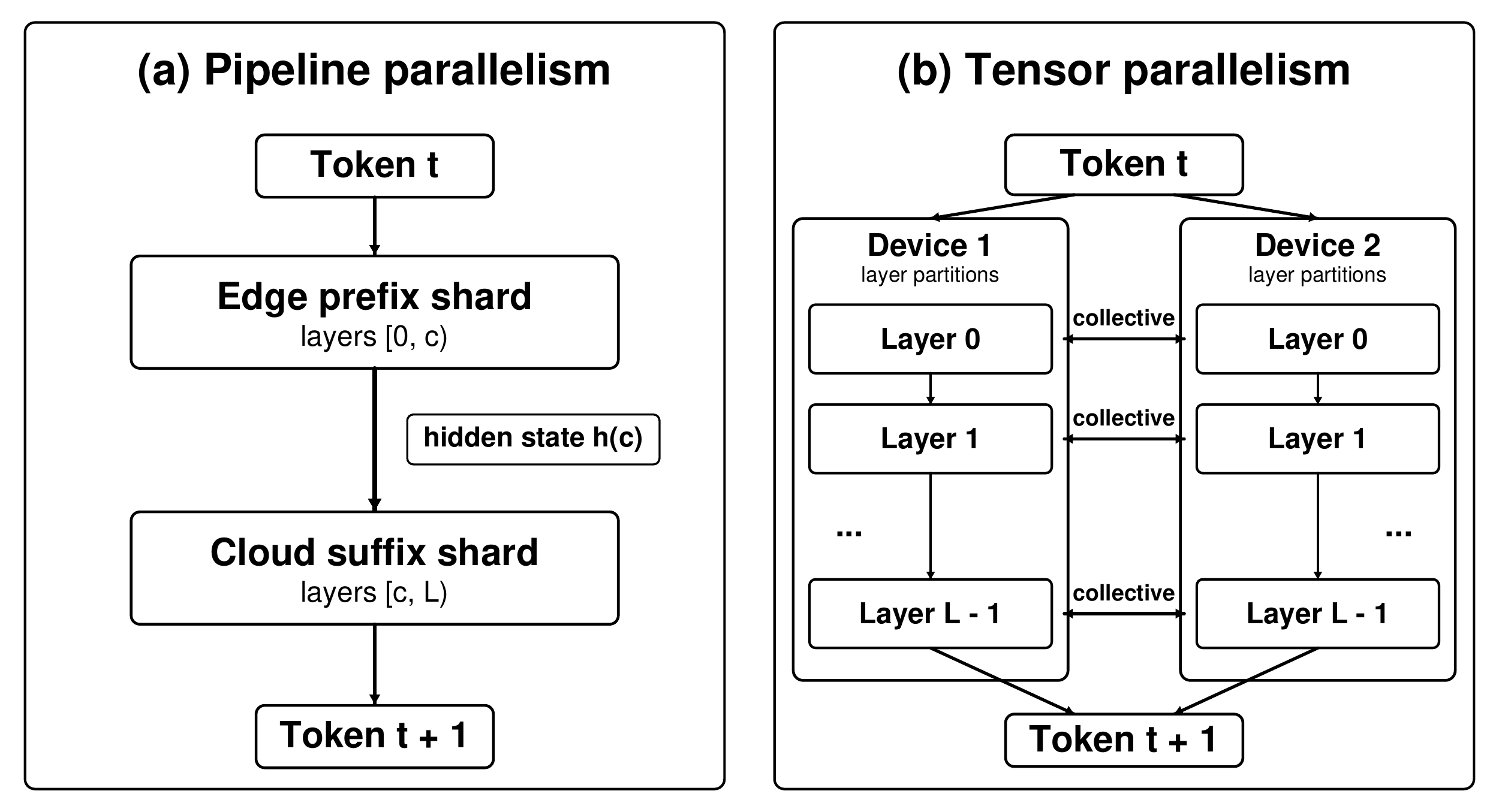}
  \caption{Pipeline and tensor parallelism for one decoding step. Pipeline sharding transfers the cut-layer hidden state, whereas tensor parallelism requires collective communication within every layer.}
  \Description{Tensor parallelism splits each model layer across devices and aggregates their outputs. Pipeline parallelism places consecutive layer ranges on different devices and transfers intermediate activations between stages.}
  \label{fig:parallelism}
\end{figure}

Distributed model execution primarily uses tensor and pipeline parallelism, as illustrated in Fig.~\ref{fig:parallelism}. Tensor parallelism divides individual operators and requires frequent collective communication, making it most suitable for tightly coupled accelerators with high-bandwidth interconnects~\cite{DeepSpeed_TP_res,EnergonAI_res}. Pipeline parallelism assigns consecutive layer ranges to different devices and exchanges the activation at each stage boundary~\cite{GPipe_res}. This communication pattern is better suited to heterogeneous edge--cloud devices, so, in this work, we adopt a pipeline-style split with one prefix and one suffix shard.

Layer-wise edge--cloud sharding introduces two forms of cloud-side 
inefficiency. Temporally, edge computation and communication create idle gaps between successive cloud invocations, which can increase the latency of the subsequent forward pass. Structurally, heterogeneous cut layers cause requests to reach the cloud at different model depths, preventing conventional batching even when they share a common suffix. We characterize the former and address the latter with depth-synchronized batching (DSB).

\begin{figure}[t]
    \centering
    \includegraphics[width=\columnwidth]{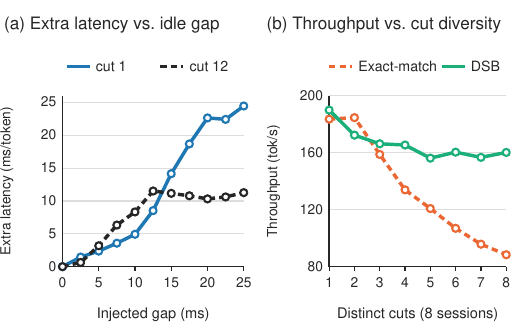}
    \caption{Motivating results from separate single-edge and multi-edge experiments. (a) Controlled idle gaps add cloud suffix latency beyond the injected wait. (b) DSB throughput is less sensitive to cut diversity than exact-match batching.}
    \label{fig:motivation}
\end{figure}

To investigate the idle-gap and cut-diversity effects illustrated in Fig.~\ref{fig:motivation}, we build DySCo, a collaborative inference runtime that keeps KV caches with their respective shards, transfers only the cut-layer hidden state and returned token during decoding, and supports different session cuts without reconstructing the model. We evaluate four Qwen and Llama configurations on heterogeneous single-edge devices and a controlled multi-edge testbed. Under Poisson arrivals with 16 warm edge workers and mean concurrency eight, DSB improves throughput by 48\% over exact-match batching while reducing mean per-session latency, showing that heterogeneous cuts can still benefit from shared cloud execution.

In summary, this paper makes the following \textit{key} contributions:

\begin{itemize}
  \item We design and implement an edge–cloud inference runtime that keeps KV caches with their respective shards. Its dyForward mechanism selects contiguous layer ranges from loaded shards, allowing different session cuts without reloading weights for each cut.

  \item We measure how cut placement affects split-inference latency across heterogeneous edge devices and network settings, and use controlled replay to isolate the additional cloud-side latency caused by inter-token idle gaps. These results show that layer count alone does not determine cloud latency.

  \item We introduce depth-synchronized batching (DSB), which advances requests with different cuts to a common model depth and executes their remaining shared suffix as one batch.

  \item We validate the prototype with four Qwen and Llama model configurations and prompts from two source datasets. Controlled replay shows up to approximately 25 ms of additional decode latency per generated token after injected waiting time is excluded; at a mean concurrency of eight, DSB improves throughput by 48\% over exact-match batching while reducing mean per-session latency.
\end{itemize}

\section{Background and Related Work}

\noindent\textbf{LLM serving and model parallelism.}
Modern LLM serving systems improve accelerator utilization through
iteration-level scheduling, continuous batching, and memory-aware KV-cache management. Orca interleaves requests at iteration granularity, vLLM introduces PagedAttention for flexible KV-cache allocation, and Sarathi-Serve combines prefill and decode work to control the throughput--latency trade-off~\cite{Orca_res,paged_res,sarathiserve2024}. Other systems distribute model execution within a datacenter. DeepSpeed Inference and EnergonAI combine model-parallel techniques for large Transformer inference, while Splitwise and DistServe place prefill and decode on separate GPU pools~\cite{DeepSpeed_TP_res, EnergonAI_res,splitwise2024,distserve2024}. These systems assume homogeneous or centrally provisioned model partitions. Our setting instead receives decode requests from heterogeneous edge devices at different layer depths, which prevents ordinary batching even when the requests share a long cloud suffix.

\noindent\textbf{Edge and hybrid LLM inference.}
On-device work reduces model size and execution cost through compact
architectures, quantization, pruning, and cache optimization; a recent survey provides a broader taxonomy of these techniques~\cite{A_Review_rw, liu2024mobilellmoptimizingsubbillionparameter_1_24}. Cloud--edge systems also offload complete requests or select among local and remote models according to resource and quality constraints~\cite{activeInference_131,Hybrid_rw}. A different line of work separates inference phases. P/D-Device moves prefill and decode between cloud and device resources~\cite{PD_Device_nBack}. These designs choose where a request or phase executes, whereas we divide every forward pass of one decoder-only model at a layer boundary and study the resulting token-level dependency. SHARD constructs nested sub-transformers from a pretrained Vision Transformer and uses incremental parameter loading to switch configurations at runtime~\cite{kalra2027shard}. Its reusable resident parameters motivate a similar systems principle, but it adapts the width of an on-device vision model; our runtime selects contiguous LLM layer ranges across devices without changing or transferring weights during inference.

\noindent\textbf{Collaborative and partitioned inference.}
Layer-wise cloud--edge partitioning predates LLMs. Neurosurgeon selects a DNN partition from device, network, and server profiles, while distributed DNNs map model sections across end, edge, and cloud resources~\cite{Neurosurgeon_dis, DDNN-dis}. For large Transformers, Petals pipelines model blocks across Internet-connected participants, EdgeShard jointly optimizes device selection and shard placement, and SplitLLM studies model placement and throughput in collaborative LLM inference~\cite{Petals-dis,edgeshard_rw,splitllm2024}. These works primarily establish capacity pooling or optimize where shards are placed. Our work is complementary in that we hold a request's cut fixed during generation, measure how repeated edge--cloud alternation changes runtime latency, and address the cloud-side batching incompatibility created when concurrent requests use different cuts. DSB exploits their common suffix rather than requiring fleet-wide agreement on one partition.

\section{Problem Setting and Overview}
\label{sec:problem-overview}

\noindent\textbf{Execution model.}
We consider a decoder-only Transformer~\cite{vaswani2017attention} with $L$ decoder layers and one or more edge sessions sharing a cloud GPU. For session $i$ with cut layer $c_i$, the edge executes the prefix $[0,c_i)$, while the cloud executes the suffix $[c_i,L)$ followed by final normalization, the LM head, and token selection. A cut is chosen at session initialization and remains fixed through prefill and decoding; different sessions may use different cuts. Each device owns the KV-cache entries for its assigned layers; KV caches are not transferred during normal inference.

Each request consists of one prefill followed by autoregressive decode. Prefill processes the complete prompt and sends the cut-layer hidden sequence to the cloud. At each decode step, the edge consumes the latest token, advances its prefix cache, and sends one cut-layer hidden state. The cloud advances its suffix cache and returns the next token ID. This repeated alternation couples the readiness and execution time of the two devices at token granularity.

\noindent\textbf{Research questions.}
This execution model exposes two complementary problems.

\emph{RQ1. How do cut placement and idle gaps jointly affect latency on heterogeneous edge--cloud devices?} A deeper cut adds additional work to edge and removes work from cloud, but affects the communication and the interval between successive GPU invocations. We decompose end-to-end latency across edge devices and use controlled hidden-state replay to isolate the effect of these idle intervals from layer computation and network transfer.

\emph{RQ2. How can a shared cloud batch sessions that arrive at different model depths?} Conventional batching applies one layer range to every row and therefore combines only sessions with identical cuts. Heterogeneous cuts fragment requests into small groups even when they share a long suffix. We address this structural incompatibility with depth-synchronized batching (DSB), which advances requests privately to the deepest cut in a scheduling window and executes their common tail as one batch.

\noindent\textbf{Approach and scope.}
DySCo uses \emph{dyForward} to select contiguous layer ranges from resident edge and cloud shards without reconstructing the model. The single-edge evaluation characterizes temporal effects across heterogeneous devices and network settings; the multi-edge evaluation isolates scheduling and batch formation under controlled hardware placement. We keep model weights unchanged, hold each request's cut fixed during generation, and leave joint cut-layer selection and scheduler adaptation to future work.

\section{System Design and Implementation}
This section presents the mechanisms that enable our collaborative edge--cloud inference system. We first describe the runtime architecture and request workflow, then introduce dynamic layer-range execution and the communication protocol, and finally describe cloud-side scheduling for multiple heterogeneous edge sessions.

\subsection{Architecture and Execution Flow}
\label{sec:architecture-flow}

\begin{figure}[t]
  \centering
  \includegraphics[width=\linewidth]{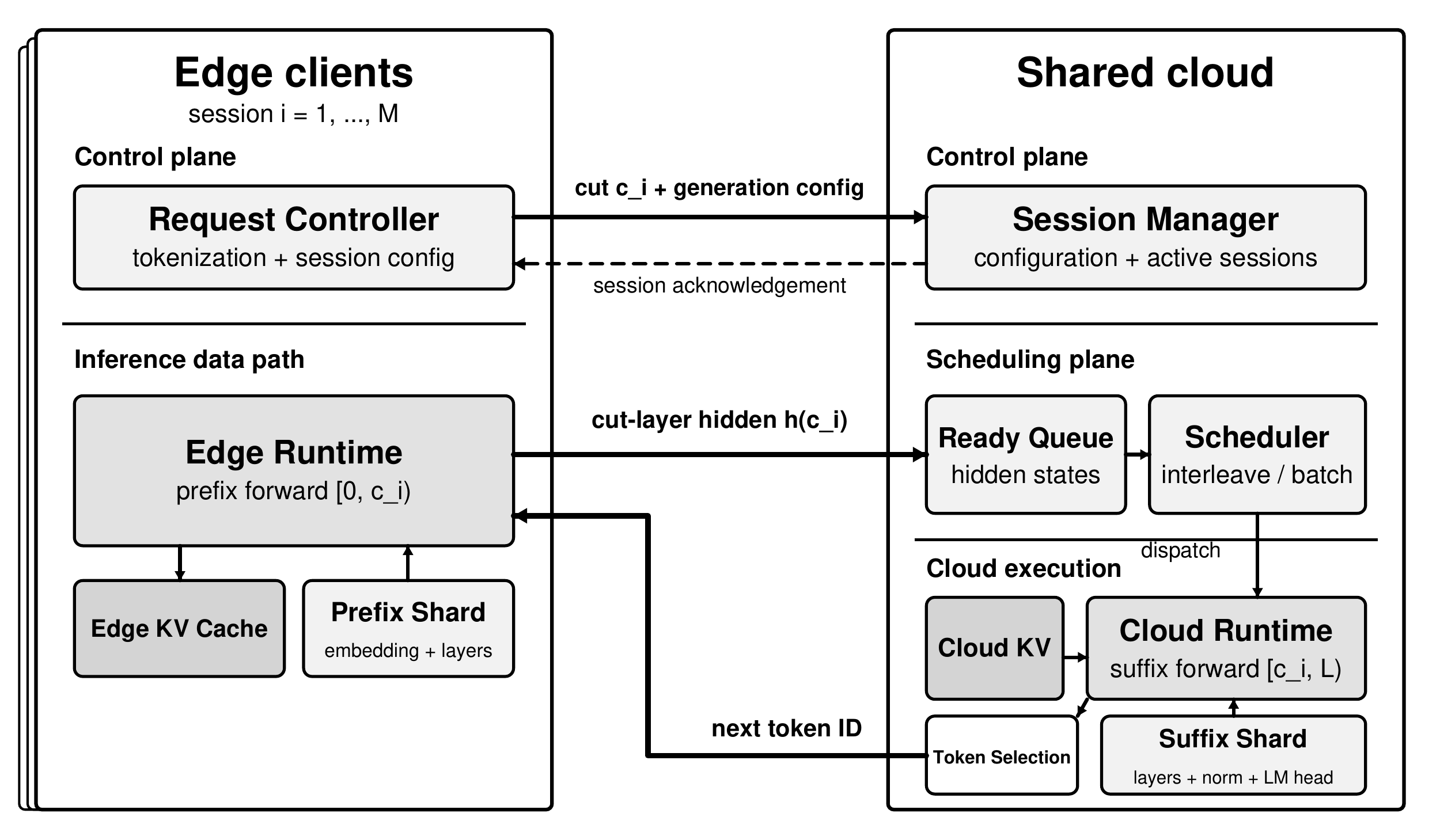}
  \caption{Edge--cloud split-inference architecture with per-device KV
  caches and a shared cloud request queue.}
  \Description{Architecture of a collaborative inference system in which multiple edge request controllers and runtimes communicate with a shared cloud session manager, scheduler, and runtime. The edge and cloud retain separate KV caches for their assigned layer ranges.}
  \label{fig:system-architecture}
\end{figure}

\begin{figure}[t]
  \centering
  \includegraphics[width=\linewidth]{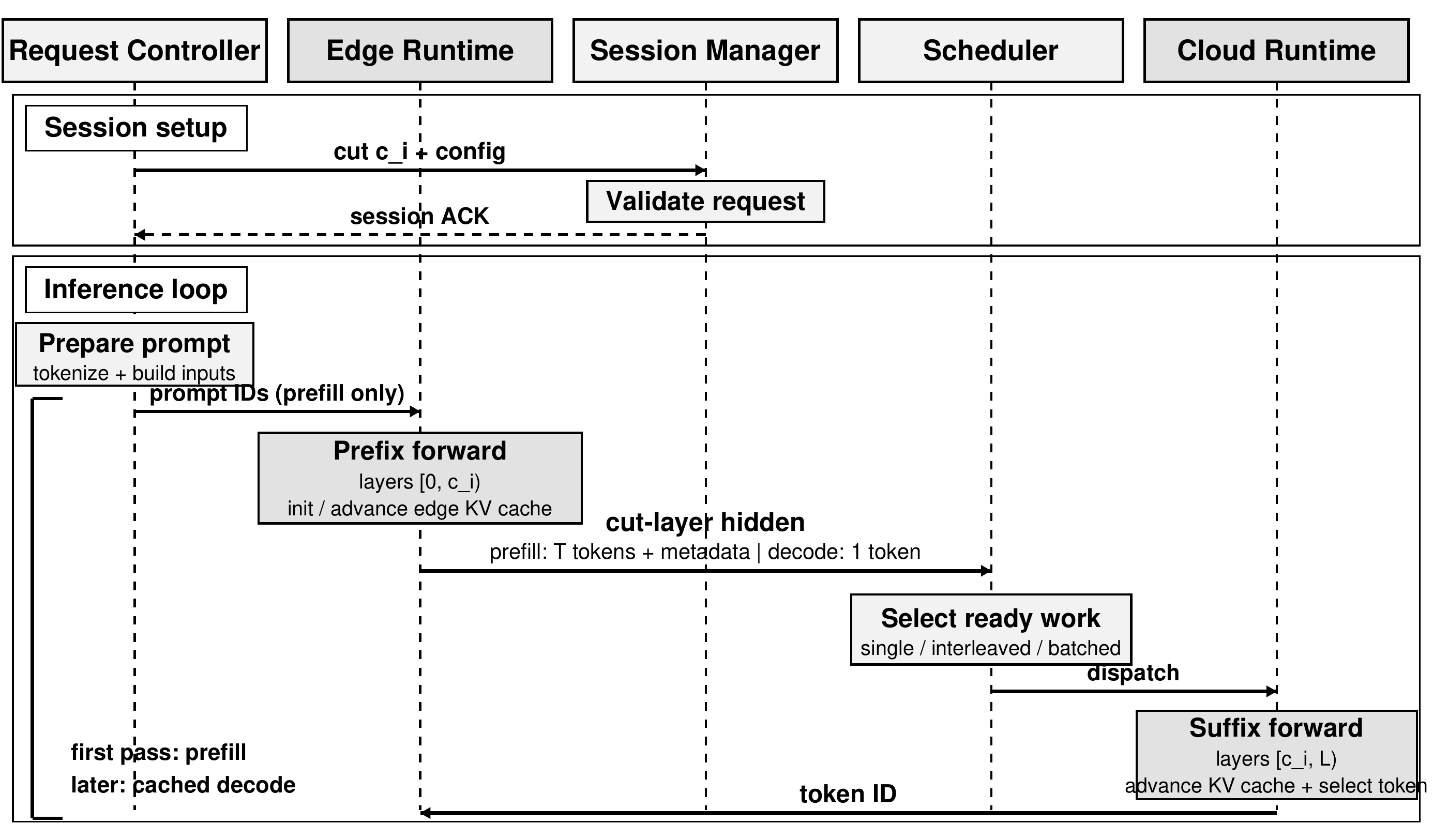}
  \caption{Session initialization, prefill, and autoregressive decode in the edge--cloud inference pipeline.}
  \Description{Sequence diagram showing session initialization followed by a shared inference loop for prefill and repeated decode steps. The cloud session manager validates the initial configuration and creates session state. The edge then executes the prefix and sends a cut-layer hidden state; the cloud scheduler dispatches the ready work, the cloud runtime executes the suffix, and the selected token ID is returned to the edge.}
  \label{fig:system-workflow}
\end{figure}

Figure~\ref{fig:system-architecture} shows the system architecture of DySCo. One or more edge clients share a cloud inference backend. Each request is associated with a cut layer $c$ in a model containing $L$ decoder layers. The edge runtime executes the prefix $[0,c)$ and maintains its prefix KV cache. The cloud runtime executes the suffix $[c,L)$, maintains the corresponding suffix KV cache, and applies the final normalization, LM head, and token-selection operation. On the edge, a request controller manages tokenization, generation configuration, and the RPC session. On the cloud, a session manager validates requests and creates per-session state, while a scheduler queues ready work
and dispatches it to the cloud runtime.

The architecture separates the control path from the repeated inference data path. During session initialization, the request controller sends the cut layer and generation configuration to the cloud session manager, which validates the request, creates the corresponding session state, and returns an acknowledgment. During inference, the edge sends the cut-layer hidden state to the cloud, while the cloud returns the selected token ID. Additional state needed to initialize or resynchronize a session is carried only when required.

Figure~\ref{fig:system-workflow} summarizes the per-request workflow. On the first pass, the request controller tokenizes the prompt and the edge runtime processes the full prompt through the prefix shard, initializing its local KV cache. The resulting cut-layer hidden states and first-pass metadata are sent to the cloud scheduler, which dispatches the ready work to the cloud runtime. The cloud executes the suffix shard, initializes its KV cache, and returns the first generated token ID.

Subsequent passes follow the same path with cached decoding: the edge consumes the latest token and advances its prefix KV cache, while the cloud advances its suffix KV cache and returns the next token ID. The scheduler may dispatch ready work individually, interleave sessions, or form a compatible batch. This loop continues until generation terminates.

KV-cache state remains local to the device that executes the corresponding layers and is not transferred on the normal per-token path. The cloud scheduler therefore operates on requests whose cut-layer hidden states are ready, dispatching them individually, interleaving sessions, or forming compatible batches without changing the edge-side execution path.

\subsection{Sharding and Communication}
\label{sec:dynamic-sharding-communication}

\begin{figure}[t]
  \centering
  \includegraphics[width=\linewidth]{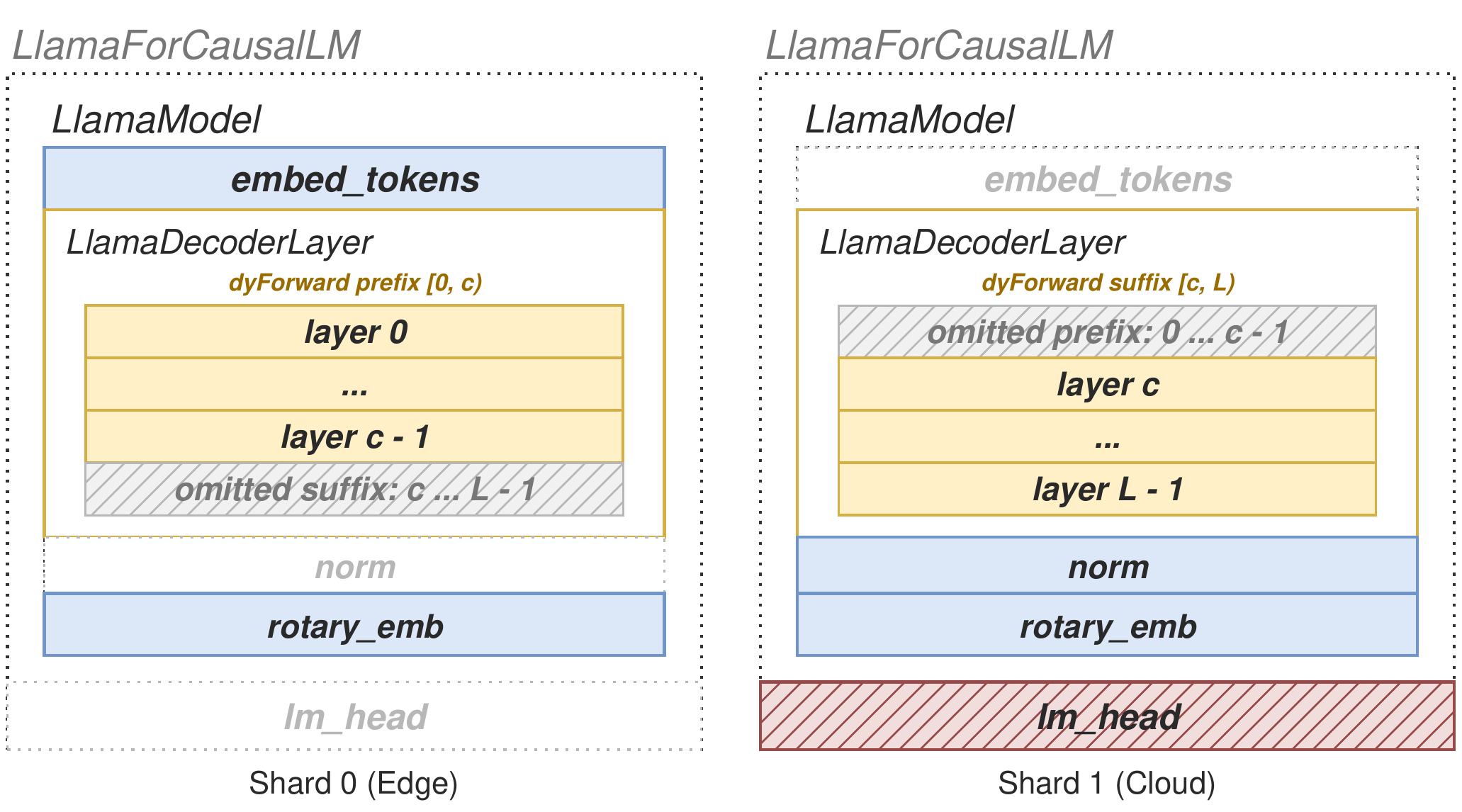}
    \caption{Dynamic layer-range execution across the resident edge and cloud shards.}

    \Description{Two side-by-side diagrams show the components resident on the edge and cloud and the decoder-layer ranges selected by dyForward.}
    \label{fig:shard-structure}
\end{figure}

\noindent\textbf{Dynamic layer-range execution (dyForward).}
For a model with $L$ decoder layers and cut layer $c$, the edge executes the prefix $[0,c)$ and the cloud executes the suffix $[c,L)$. The edge shard contains the token embedding and the decoder layers that it may execute, whereas the cloud retains the decoder layers required for suffix execution, final normalization, LM head, and token-selection logic. Each side maintains KV-cache entries only for the layers it executes.

Materializing a separate model instance for every possible cut would duplicate weights and require expensive reconstruction whenever the partition changes. To avoid this, we implement \textit{dyForward}, a model-aware layer-range executor for Llama and Qwen that dynamically selects a contiguous decoder range from a resident model shard. Given an input hidden state, a start layer, and an optional end layer, dyForward invokes only the selected decoder layers. It constructs the model-specific causal mask and positional inputs, updates the corresponding KV cache, and applies final normalization only when execution reaches the end of the model. A participating edge holds a prefix shard up to its supported depth, and \textit{dyForward} selects the active layers when each session is initialized. Different sessions can therefore assign different amounts of computation to the edge while reusing resident edge weights and a shared cloud model instance, without moving weights for each cut. The cut remains fixed during generation, so each layer's KV cache stays on the device executing that layer. Figure~\ref{fig:shard-structure} summarizes this organization.

\begin{table}[t]
  \centering
  \caption{Messages transmitted during normal execution.}
  \label{tab:payload}
  \small
  \begingroup
  \setlength{\tabcolsep}{3pt}
  \renewcommand{\arraystretch}{1.08}
  \begin{tabular}{@{}l
    >{\raggedright\arraybackslash}p{0.37\columnwidth}
    >{\raggedright\arraybackslash}p{0.28\columnwidth}@{}}
    \toprule
    \textbf{Phase} & \textbf{Edge $\rightarrow$ Cloud}
      & \textbf{Cloud $\rightarrow$ Edge} \\
    \midrule
    Setup & Cut layer and generation configuration & Session acknowledgment \\
    Prefill & Hidden states, attention mask, and position IDs & Token ID \\
    Decode & Hidden state & Token ID \\
    \bottomrule
  \end{tabular}
  \endgroup
\end{table}

\noindent\textbf{Normal-path communication.}
Table~\ref{tab:payload} lists the messages exchanged during normal execution. The cut layer and generation configuration are fixed when a session is created. Prefill sends the cut-layer hidden sequence together with the attention mask and position IDs required to initialize the cloud state. During cached decoding, both devices derive the mask and current position from their local session state; the repeated cross-device path therefore contains only one hidden state in the edge-to-cloud direction and one token ID in the reverse direction. KV caches and intermediate layer state remain local. Recovery, migration, or explicit resynchronization may exchange additional state, but such exceptional traffic is outside the normal inference path and is omitted from the table.

The principal tensor sizes can be estimated analytically:
\[
\begin{aligned}
S_{\mathrm{hidden}}
  &= B T_{\mathrm{in}} d_{\mathrm{model}} b_h, \\
S_{\mathrm{pos}}
  &= B T_{\mathrm{in}} b_{\mathrm{pos}},
  \qquad
S_{\mathrm{mask}}
  = B T_{\mathrm{ctx}} b_{\mathrm{mask}}.
\end{aligned}
\]
Here, $B$ is the batch size, $T_{\mathrm{in}}$ is the number of tokens processed in the current forward call, and $T_{\mathrm{ctx}}$ is the total context length represented by the current request state. The terms $b_h$, $b_{\mathrm{pos}}$, and $b_{\mathrm{mask}}$ denote the corresponding bytes per element.

During prefill, $T_{\mathrm{in}}=T_{\mathrm{ctx}}$ equals the prompt length. During cached decoding, $T_{\mathrm{in}}=1$, while $T_{\mathrm{ctx}}$ includes the prompt and previously generated tokens. Because the mask and position IDs are not retransmitted during normal decode, the recurring payload is dominated by the single-token hidden state. Session configuration contains only scalar control values and contributes negligible per-token cost.

\noindent\textbf{Transport.}
Cross-device communication uses gRPC. Tensor serialization, RPC framework overhead, and network transfer are all included in the end-to-end measurements reported in the evaluation.

\subsection{Cloud-Side Scheduling}
\label{sec:cloud-scheduling}

\begin{figure}[t]
  \centering
  \includegraphics[width=0.9\linewidth]{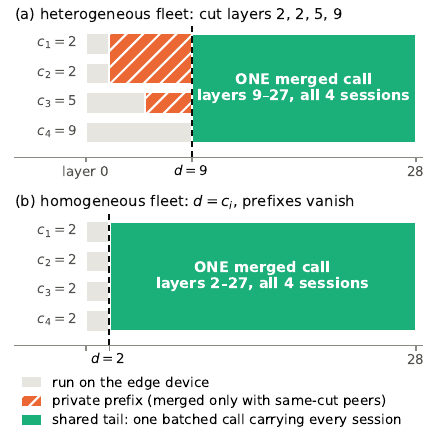}
  \caption{Depth-synchronized batching. Sessions with different cut layers advance over private ranges to $d=\max_i c_i$ and then share one batched forward call over $[d,L)$. For equal cut layers, DSB reduces to exact-match batching.}
  \Description{Two diagrams compare a heterogeneous scheduling window, in which sessions advance privately to the deepest cut layer before sharing a batched suffix, with a homogeneous window that directly uses exact-match batching.}
  \label{fig:me-method}
\end{figure}

\begin{algorithm}[t]
\caption{Depth-synchronized decode step}
\label{alg:dsb}
\begin{algorithmic}[1]
\REQUIRE ready window $W$; cut layers $c_i$; model depth $L$
\STATE $d \gets \max_{i\in W} c_i$
\FORALL{distinct $c<d$ represented in $W$}
    \STATE $G_c \gets \{i\in W:c_i=c\}$
    \STATE $h_i \gets \textsc{RunRange}(G_c,[c,d))$ for $i\in G_c$
\ENDFOR
\FORALL{$i\in W$ with $c_i=d$}
    \STATE $h_i \gets$ hidden state received from edge $i$
\ENDFOR
\STATE $(H,K,M) \gets \textsc{Merge}(\{h_i,K_i\}_{i\in W})$
\STATE $(Z,K') \gets \textsc{RunRange}((H,K),[d,L))$
\STATE $\{K_i'\}_{i\in W} \gets \textsc{Split}(K',M)$
\RETURN $\textsc{SelectToken}(Z_i)$ for each $i\in W$
\end{algorithmic}
\end{algorithm}

Because sessions can be initialized with different cuts from their resident edge shards, a shared cloud model receives decode requests at different layers. A conventional batched forward call applies the same layer range to every row, so exact-match batching can merge only sessions with the same cut layer. As device heterogeneity increases, the cloud therefore executes more small forward calls and repeatedly reads the same suffix weights.

\noindent\textbf{Depth-synchronized batching (DSB).}
DSB recovers batching across sessions with different cut layers. Let $W$ be a scheduling window, $c_i$ the cut layer of session $i\in W$, and $L$ the model depth. DSB chooses the deepest arrival layer as a synchronization checkpoint,
\[
d=\max_{i\in W} c_i.
\]
Each session first advances over its private range $[c_i,d)$. Sessions sharing the same $c_i$ may already be batched for this range. Once all sessions reach $d$, their hidden states and suffix KV-cache rows are merged and one batched forward call executes the common tail $[d,L)$. The resulting cache rows and token outputs are then returned to their corresponding sessions.

Figure~\ref{fig:me-method} contrasts heterogeneous and homogeneous windows. The checkpoint must equal the deepest cut because layers before $c_i$ have already executed on edge $i$, while every session still requires the complete suffix from its own cut to $L$. When all cuts coincide, $d=c_i$, the private ranges vanish and DSB reduces exactly to exact-match batching. A device joining or leaving changes only the composition and checkpoint of a subsequent scheduling window; no fleet-wide cut-layer coordination is required.

Algorithm~\ref{alg:dsb} summarizes one DSB decode step. The scheduler operates only on ready requests collected in the current window. KV caches remain session-local outside a merged call; padding and row maps temporarily reconcile different context lengths, after which the updated rows are split back into the per-session states. Prefill remains unbatched in the current implementation, so DSB changes only cloud-side decode dispatch and does not alter the edge execution path.

\FloatBarrier
\section{Performance Evaluation}

We evaluate DySCo at two complementary scales. The single-edge experiments characterize how partition depth, device heterogeneity, and inter-token idle gaps affect split inference. The multi-edge experiments then evaluate whether depth-synchronized batching (DSB) can recover cloud throughput when concurrent sessions enter the cloud at different layers.

\subsection{Experimental Setup}

\textbf{Testbeds and scale settings.}
The single-edge testbed varies the edge hardware while keeping the cloud GPU fixed, allowing us to characterize a wide range of edge-side compute capabilities. It combines a low-power Jetson platform with several datacenter GPUs, summarized in Table~\ref{tab:single-edge-devices}. An RTX A6000 is used as the default cloud GPU because it provides sufficient memory for the complete model and long-context profiling.
Communication uses gRPC between separate edge and cloud processes. Datacenter GPU pairs communicate over a private intra-cluster network, whereas the remote Jetson--cloud pair communicates over a WAN through SSH port forwarding. 

In contrast, the multi-edge testbed fixes the hardware and varies the number and cut layers of concurrent sessions. Each edge process executes a front shard and connects over gRPC to a single cloud process, which holds one model copy and serves all sessions regardless of their cut layers. The processes run on one multi-GPU server (cloud: RTX PRO 6000 Blackwell, 96\,GB; edges: RTX PRO 4500 Blackwell, 32\,GB), isolating scheduling behavior from wide-area network variation. All multi-edge processes run in the same containerized environment.

\begin{table}[t]
\centering
\small
\caption{Devices used in the single-edge evaluation.}
\label{tab:single-edge-devices}
\begin{tabular}{l|l|l|l}
\hline
\textbf{Device} & \textbf{GPU mem.} & \textbf{CPU} & \textbf{RAM} \\
\hline
Jetson Orin & Shared & Cortex-A78AE & 8 GB \\
RTX A2 & 16 GB & AMD EPYC 7402P & 128 GB \\
RTX A5000 & 24 GB & AMD EPYC 7402P & 128 GB \\
RTX A6000 & 48 GB & 2 $\times$ AMD EPYC 7402 & 128 GB \\
\hline
\end{tabular}
\end{table}

\textbf{Models and workloads.}
At each scale, we evaluate one Qwen model and one Llama model. The single-edge study uses Qwen3.5-9B~\cite{qwen2026qwen35} (32 layers) for its main results and Llama-3.1-8B~\cite{grattafiori2024llama3} (32 layers) for replication. The multi-edge study uses Llama-3.2-3B~\cite{meta2024llama32} (28 layers) and replicates its main trend with Qwen2.5-3B~\cite{yang2024qwen25} (36 layers).

Both studies draw from the same prompt pools. WildGPT-4, the GPT-4 subset of WildChat~\cite{zhao2024wildchat}, supplies short prompts (16--128 tokens) for single-edge decode, controlled replay, and the multi-edge scheduling sweeps. LongReason~\cite{ling2025longreason} supplies long prompts (approximately 4,000--6,000 tokens) for single-edge prefill and the multi-edge context-length sweep, where prompts are truncated to the target lengths. Prompts are tokenized separately for each model, and we retain subsets with comparable tokenized lengths. Measured prompt and session counts are reported with the corresponding results.

\textbf{Execution protocol.}
Unless stated otherwise, inference uses bfloat16 and greedy decoding, with thinking disabled for Qwen. Single-edge requests generate exactly 64 tokens, whereas multi-edge sessions generate 128 tokens; early stopping is disabled in both settings. Across both evaluation scales, the runtime uses \textit{dyForward} to vary the cut layer without reloading model weights. Each session keeps its assigned cut from prefill through decoding; cuts vary across sessions and experimental conditions. For cut layer $c$, the edge executes layers $[0,c)$ and the cloud executes $[c,L)$. In the multi-edge evaluation, we emulate heterogeneous edge capacities by assigning different cut layers while keeping the physical edge GPUs identical. For Llama-3.2-3B in bfloat16, the edge footprint consists of a 788 MB embedding table plus approximately 201 MB per decoder layer, so smaller emulated memory budgets correspond to shallower cuts.

\subsection{Results and Analysis}

\subsubsection{Single-Edge Single-Cloud Characterization}
The single-edge study is intended to characterize latency rather than demonstrate throughput gains, since a single request cannot fully utilize a powerful cloud GPU. Instead, it establishes how a cut changes latency on both devices and identifies the constraints that a multi-edge scheduler must account for.

\begin{figure}[t]
    \centering
    \begin{subfigure}[t]{0.485\linewidth}
        \centering
        \includegraphics[width=\linewidth]
        {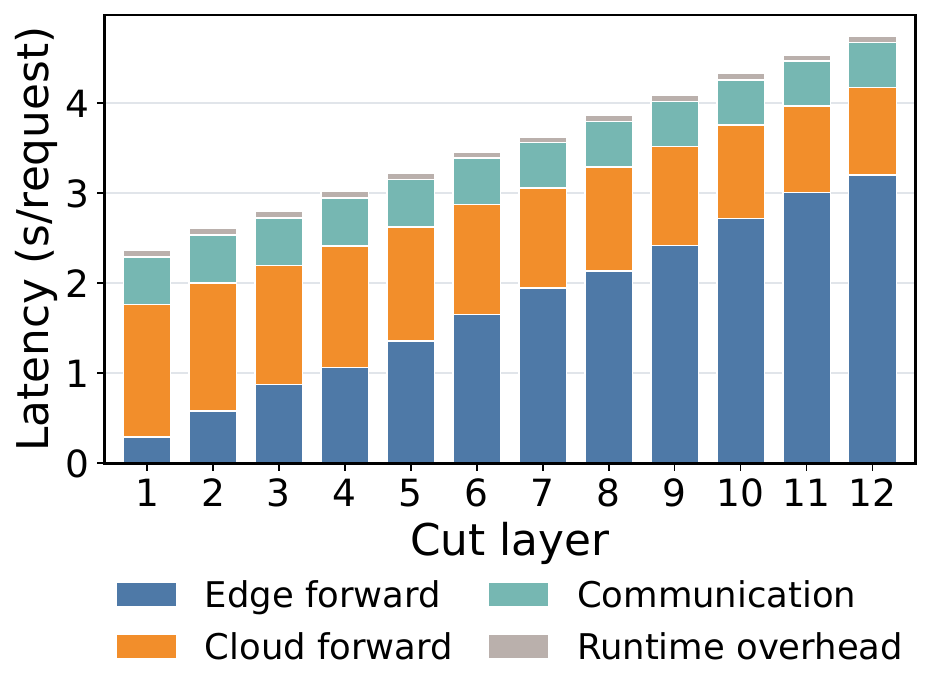}
        \caption{Long-input prefill.}
        \label{fig:a2-prefill-breakdown}
    \end{subfigure}
    \hfill
    \begin{subfigure}[t]{0.485\linewidth}
        \centering
        \includegraphics[width=\linewidth]
        {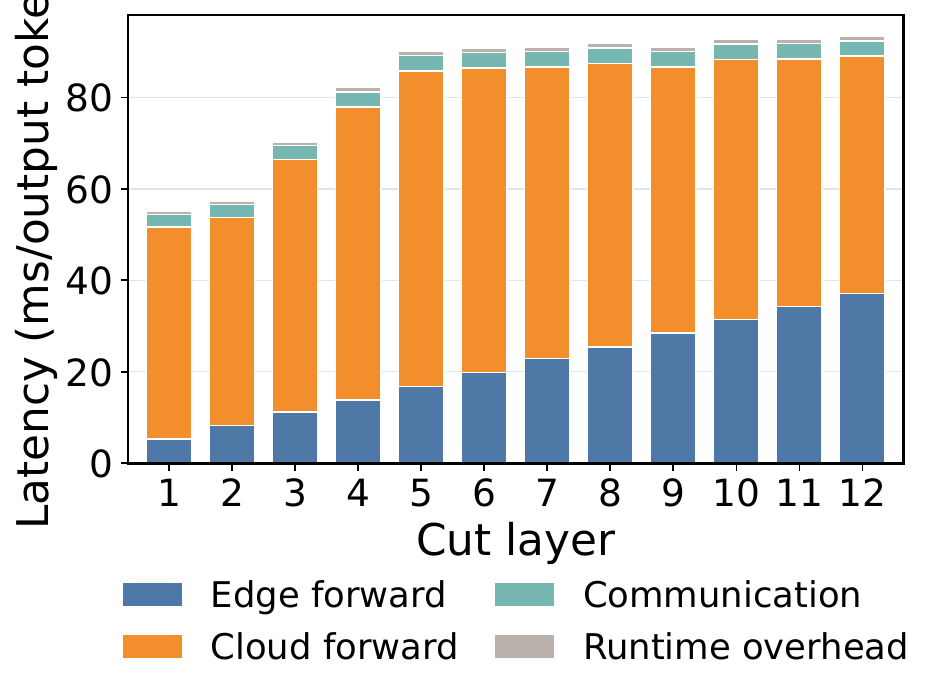}
        \caption{Short-input decode.}
        \label{fig:a2-decode-breakdown}
    \end{subfigure}
    \caption{Mean latency breakdowns for Qwen3.5-9B with an A2 edge
    and an A6000 cloud.}
    \label{fig:a2-a6000-breakdown}
\end{figure}

\textbf{Latency decomposition.}
For each cut layer, we execute 80 short and 10 long inputs and exclude the first input of each group after a cut change as warm-up. Figure~\ref{fig:a2-a6000-breakdown} presents two representative views of an A2--A6000 deployment. Long-input prefill highlights the cost of processing the full prompt and transmitting its cut-layer activations, whereas short-input decode captures the repeated per-token execution path that determines steady-state decoding latency. Prefill communication is larger because the cut-layer activation contains the complete input sequence; cached decode transfers only the activation of the new token. We therefore retain prefill as a validation case and focus the remaining characterization on short-input decode.

\begin{figure}[t]
    \centering
    \begin{subfigure}[t]{0.485\linewidth}
        \centering
        \includegraphics[width=\linewidth]
        {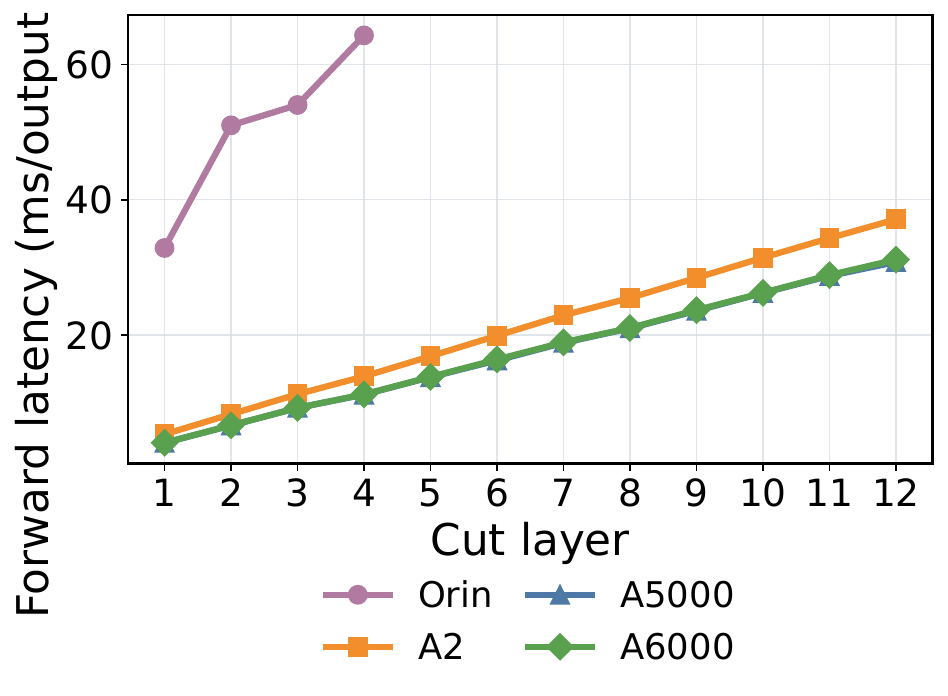}
        \caption{Edge forward latency.}
        \label{fig:edge-forward-devices}
    \end{subfigure}
    \hfill
    \begin{subfigure}[t]{0.485\linewidth}
        \centering
        \includegraphics[width=\linewidth]
        {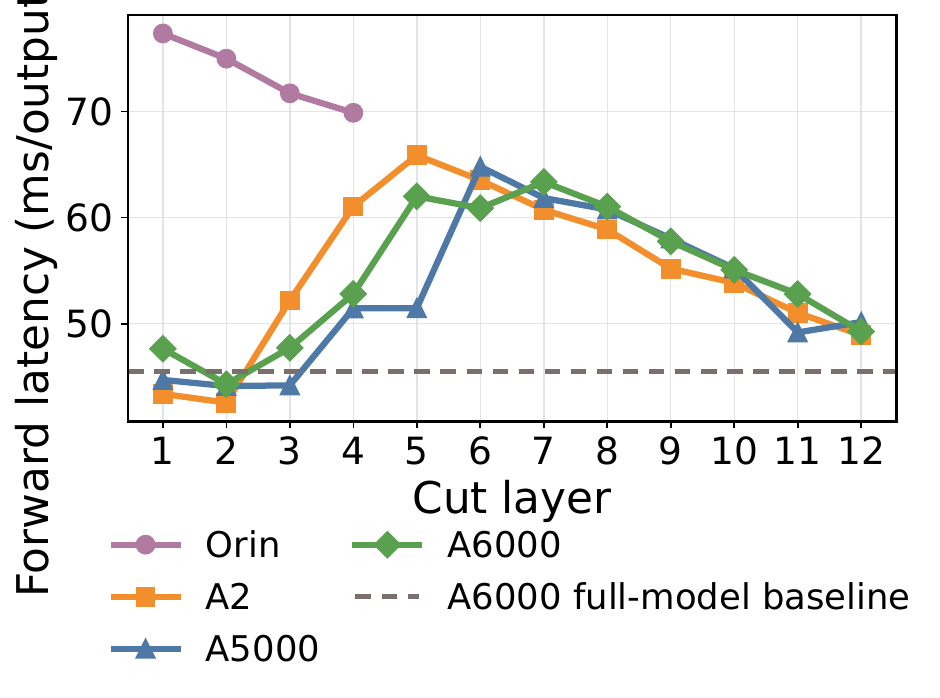}
        \caption{Cloud forward latency.}
        \label{fig:cloud-forward-devices}
    \end{subfigure}
    \caption{Short-input decode latency across edge devices with the
    cloud fixed to the same A6000.}
    \label{fig:forward-across-edge-devices}
\end{figure}

\textbf{Cut-layer behavior across edge devices.}
The Jetson Orin can hold only four decoder layers, so we additionally use an A2, A5000, and a second A6000 as edge devices while fixing the cloud to the same A6000. Figure~\ref{fig:forward-across-edge-devices} shows that edge forward latency initially appears to increase approximately linearly with the number of prefix layers. Cloud latency, however, does not simply decrease with the remaining suffix depth. Except for the Orin, it first increases sharply, reaches a device-dependent turning point, and only then decreases as layers are removed from the cloud. It can consequently exceed the A6000 full-model baseline even though the cloud executes fewer layers. The same qualitative behavior appears with Llama-3.1-8B, indicating that it is not specific to the Qwen architecture.

After the turning point, the cloud curves follow a similar decreasing trajectory. This behavior suggests that cloud execution depends not only on suffix depth but also on the idle interval while the edge computes and transmits the next hidden state. The Orin is sufficiently slow that the cloud curve begins directly in this high-overhead regime.

\textbf{Controlled idle-gap experiment.}
To isolate this effect, we replay pre-generated cut-layer hidden states on the A6000, removing edge computation and network variation while preserving the same suffix execution. For cut layers 1--12, we insert controlled gaps of 0--25\,ms between decode invocations. Each configuration uses 32 short inputs; the first is warm-up and the remaining 31 are measured. We exclude the injected waiting time and the cut-specific zero-gap reference:
\[
\Delta L(c,d)=L(c,d)-d-L(c,0),
\]
where $c$ is the cut layer and $d$ the injected gap. To reduce measurement noise, $L(c,0)$ is estimated from a smoothed zero-gap baseline; non-zero-gap measurements are not smoothed.

\begin{figure}[t]
    \centering
    \includegraphics[width=\linewidth]
    {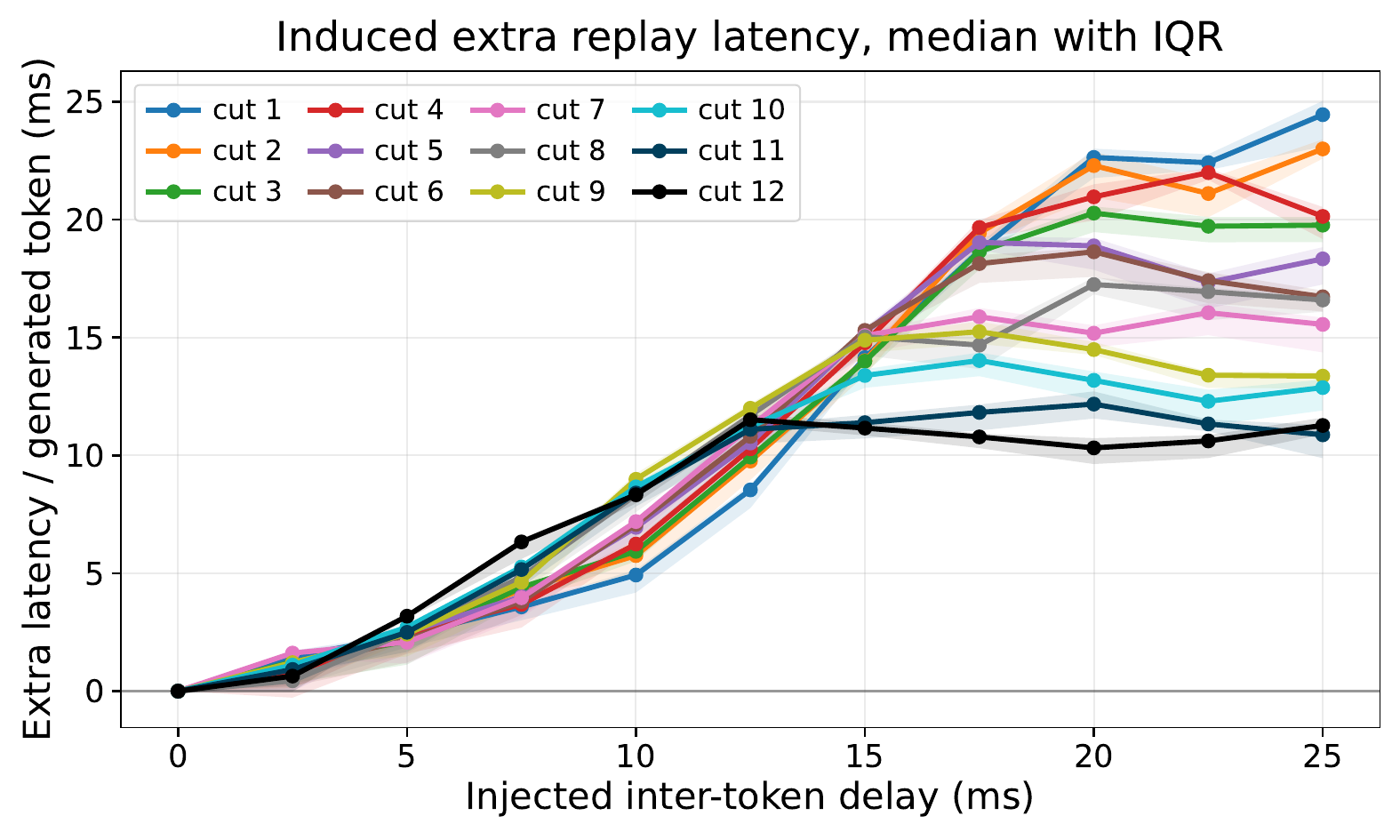}
    \caption{Additional A6000 decode latency after controlled idle gaps.
    Lines show the median over 31 prompts; shaded regions span the 25th--75th
    percentiles.}
    \label{fig:suffix-idle-gap}
\end{figure}

Figure~\ref{fig:suffix-idle-gap} shows that every tested non-zero gap introduces additional execution latency beyond the waiting time itself. This overhead is not proportional to the gap. Instead, it increases over a relatively narrow range and then reaches a cut-dependent ceiling, beyond which it remains approximately constant. A larger cloud suffix has a higher ceiling and generally requires a longer gap to reach it. The maximum observed overhead is approximately 25\,ms at cut layer 1 and 13\,ms at cut layer 12. Longer exploratory measurements show little further change beyond approximately 25\,ms. A separate prefix-only experiment shows the same qualitative effect on edge execution, although its absolute cost is smaller when only a few layers are assigned to the edge.

This behavior explains the non-linearity in the end-to-end experiments. An idle gap introduces overhead immediately, but cloud latency exceeds the full-model baseline only when this overhead becomes larger than the computation saved by moving layers to the edge. Before the ceiling is reached, the overhead may grow faster than the suffix workload decreases, causing cloud latency to rise. After reaching the ceiling, moving additional layers to the edge again reduces cloud latency. Thus, layer count alone is insufficient to predict split-inference latency.

\subsubsection{Multi-Edge Serving with Depth-Synchronized Batching}

We evaluate the DSB scheduler introduced in Section~\ref{sec:cloud-scheduling} under concurrent sessions with heterogeneous cut layers. The experiments compare throughput and per-session latency while varying offered load, shared-tail depth, cut-layer diversity, fleet composition, context length, and concurrency. 

\textbf{Compared schedulers.} All schedulers use the same edge/cloud split and model instance. The GPU executes one forward call at a time under a single lock, and each prompt is processed in one unbatched prefill call. The schedulers therefore differ only in how decode steps are dispatched and combined. 

\begin{itemize}
\item \emph{FIFO} serves as the serialized reference. Each request is handled by its own server thread, which executes the session's layers $[c_i, L)$ as a single-session forward call once it acquires the GPU lock. Sessions never share a call, and no dispatcher coordinates them.

\item \emph{Exact-match batching} introduces a dedicated dispatcher thread. Decode requests are queued for up to 8\,ms, with at most eight requests per window; the dispatcher groups them by cut layer and executes each group as one batched forward call over $[c, L)$, left-padding the key/value caches of its members to a common length and splitting them afterwards. Sessions with different cut layers are executed in separate calls.

\item \emph{Round-robin interleaving} never merges sessions. A single worker cycles through a fixed group of admitted sessions (eight in the burst experiments and sixteen under Poisson arrivals) and serves one session per forward call. If the scheduled session has not yet delivered its hidden state, the worker serves the next ready session instead of idling, and sessions beyond the group size wait in a first-come overflow queue. Interleaving thus removes idle time between single-session calls but does not reduce the number of calls.

\item \emph{DSB} uses the same 8\,ms dispatch window and maximum batch size of eight as exact-match batching, but combines heterogeneous sessions through the depth-synchronization procedure in Section~\ref{sec:cloud-scheduling}.
\end{itemize}

\textbf{Measurement protocol.} The order of conditions within each measurement point and the order of points within each sweep are randomized, so that drift over a run contributes noise rather than bias. We use Welch's $|t| > 2.5$ as our significance criterion. Two controls bound the resolution of six-trial measurements. Near-identical fleets measured in separate sweeps differ by three percentage points, and a fleet with a single cut layer, for which both schedules are provably identical, yields $+3.4\%$ rather than zero. Differences below approximately six percentage points are therefore not interpreted.

\textbf{Exact-match batching under heterogeneity.} Before comparing DSB, we quantify the benefit and limitation of exact-match batching relative to FIFO. With $N{=}8$ sessions on Llama-3.2-3B, it improves throughput by $+377.0\%$ when all sessions cut at layer 2, by $+180.3\%$ for a two-tier fleet, and by $+80.9\%$ when all eight cut layers differ. In the last configuration no two sessions can be merged; the remaining gain comes from consolidating dispatch in one scheduling thread. Exact-match batching therefore loses its principal benefit as cut-layer diversity increases.

\textbf{Performance under Poisson arrivals.} Table~\ref{tab:me-poisson} compares the four schedulers under Poisson arrivals against a pool of 16 warm edge workers with cut layers 2 to 9; arrival rates are set to target an average concurrency $\bar{N}$ according to Little's law. DSB achieves the lowest mean latency at every load (4.93 to 6.45\,s, compared with 26.56 to 28.34\,s for FIFO) and the highest throughput from $\bar{N}{=}4$ upward, with statistically significant improvements on both metrics. At $\bar{N}{=}2$, its throughput advantage is within noise. At $\bar{N}{=}8$, DSB improves throughput by $+275\%$ over FIFO, $+48\%$ over exact-match batching ($t = 26.9$), and $+79\%$ over interleaving. FIFO throughput remains near 59\,tok/s while its latency grows, as a serialized scheduler saturates immediately under continuous arrivals. The throughput gain of DSB is therefore not obtained at the expense of queueing delay.

\begin{table}[t]
\centering
\caption{Throughput and mean per-session latency under Poisson arrivals at mean concurrency $\bar{N}$ (Llama-3.2-3B, WildGPT-4, 128 generated tokens; 16 workers at cut layers
2--9; 15 trials of 24 sessions). Percentages are relative to FIFO; bold
denotes a significant best result ($|t|>2.5$).}
\label{tab:me-poisson}
\begin{tabular}{c|cccc}
\hline
\textbf{$\bar{N}$} & \textbf{FIFO} & \textbf{Exact-match} & \textbf{Interleave} & \textbf{DSB} \\
\hline
\multicolumn{5}{c}{\emph{throughput} (tok/s)} \\
\hline
2 & 58.3 & 122.6 (+110\%) & 119.0 (+104\%) & 129.8 (+122\%) \\
4 & 58.8 & 142.6 (+143\%) & 121.4 (+107\%) & \textbf{186.3 (+217\%)} \\
8 & 58.5 & 148.3 (+153\%) & 122.5 (+109\%) & \textbf{219.8 (+275\%)} \\
\hline
\multicolumn{5}{c}{\emph{mean latency} (s)} \\
\hline
2 & 26.56 & 7.94 & 7.20 & \textbf{4.93} \\
4 & 27.52 & 10.12 & 11.99 & \textbf{5.84} \\
8 & 28.34 & 10.46 & 13.16 & \textbf{6.45} \\
\hline
\end{tabular}
\end{table}

\begin{figure}[t]
\centering
\includegraphics[width=\linewidth]{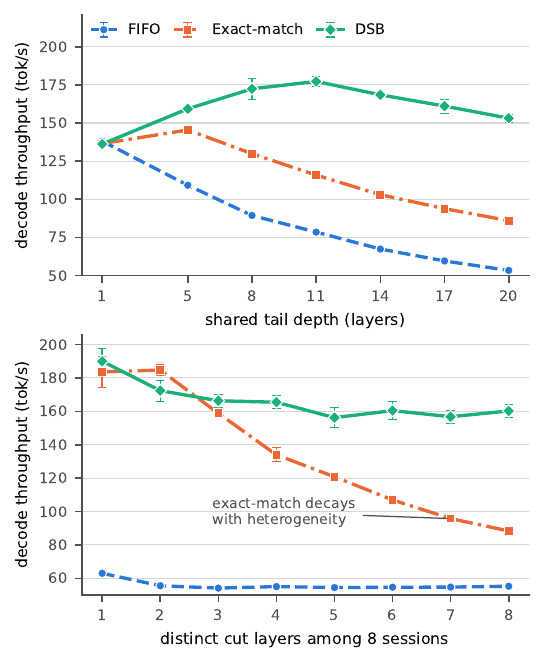}
\caption{DSB and exact-match throughput versus shared-tail depth (top)
and cut-layer diversity (bottom) for eight Llama-3.2-3B sessions with
WildGPT-4 prompts. Each sweep fixes the other variable; error bars show
one standard deviation over six trials.}
\label{fig:me-mechanism}
\end{figure}

\textbf{Determinants of the throughput gain.} Two variables govern the gain, and Fig.~\ref{fig:me-mechanism} isolates each. The first is the \emph{shared tail depth} $L - d$. With eight consecutive distinct cut layers shifted from shallow to deep positions, the gain over exact-match batching decreases monotonically from $+78.1\%$ at a 20-layer tail to $-0.2\%$ at a single layer ($r = 0.981$, $+4.48\%$ per layer). Across this sweep, DSB throughput varies by $30\%$, compared with $69\%$ for exact-match batching, so DSB substantially reduces the sensitivity of throughput to cut-layer placement. The same sweep on Qwen2.5-3B at matched relative depths (Table~\ref{tab:me-models}) reproduces this trend ($r = 0.950$, up to $+66.0\%$). Its per-layer slope is lower ($+3.18\%$ versus $+4.48\%$), close to the ratio of $28/36$ expected because each Qwen layer constitutes a smaller fraction of the model, and its gain begins slightly deeper. No gain is observed at a 17\% tail, where Llama already gains $+9.5\%$. The second variable is \emph{cut-layer diversity}. With the tail fixed and eight sessions distributed over $k$ distinct cut layers, the gain is $-6.7\%$ at $k{=}2$ and then increases linearly ($r = 0.995$, $+14.6\%$ per additional cut layer) to $+23.5\%$ at $k{=}4$, $+29.3\%$ at $k{=}5$, and $+81.3\%$ at $k{=}8$. A fleet with five distinct device classes thus already obtains a substantial gain, so the benefit is not confined to fully spread fleets. In contrast, the reduction in forward calls does not predict the gain. For example, a fleet of seven sessions at layer 10 and one at layer 2 requires $41\%$ fewer calls yet gains only $+6.3\%$. Finally, because $d$ is the deepest cut layer, the benefit is determined by the deepest device in the window rather than by a typical one.

\begin{table}[t]
\centering
\footnotesize
\setlength{\tabcolsep}{4pt}
\caption{DSB throughput gain over exact-match batching for eight
consecutive cut layers at matched relative tail depths on both models
(WildGPT-4, 128 generated tokens; six trials). Parentheses give tail depth
as a fraction of model depth; bold denotes $|t|>2.5$.}
\label{tab:me-models}
\begin{tabular}{cc|cc}
\hline
\multicolumn{2}{c|}{\textbf{Llama-3.2-3B} (28 layers)} & \multicolumn{2}{c}{\textbf{Qwen2.5-3B} (36 layers)} \\
\textbf{Tail} & \textbf{Gain} & \textbf{Tail} & \textbf{Gain} \\
\hline
20 (71\%) & $\mathbf{+78.1\%}$ & 26 (72\%) & $\mathbf{+66.0\%}$ \\
17 (61\%) & $\mathbf{+71.6\%}$ & 22 (61\%) & $\mathbf{+64.9\%}$ \\
14 (50\%) & $\mathbf{+63.5\%}$ & 18 (50\%) & $\mathbf{+55.4\%}$ \\
11 (39\%) & $\mathbf{+53.0\%}$ & 14 (39\%) & $\mathbf{+51.9\%}$ \\
8 (29\%)  & $\mathbf{+32.7\%}$ & 10 (28\%) & $\mathbf{+20.7\%}$ \\
5 (18\%)  & $\mathbf{+9.5\%}$  & 6 (17\%)  & $-0.7\%$ \\
1 (4\%)   & $-0.2\%$           & 1 (3\%)   & $-0.9\%$ \\
\hline
\multicolumn{2}{c|}{$r = 0.981$, $+4.48\%$ per layer} & \multicolumn{2}{c}{$r = 0.950$, $+3.18\%$ per layer} \\
\hline
\end{tabular}
\end{table}

\textbf{Sensitivity to fleet composition.} DSB can reduce throughput on clustered fleets (Table~\ref{tab:me-shape}). Among 17 fleets of eight sessions with at most three distinct cut layers, the eight fleets in which a single session deviates from the majority all gain over exact-match batching ($+2.2\%$ to $+16.6\%$; four significantly), whereas the six fleets whose minority comprises two or more sessions (a pair, additional clusters, or two separate outliers) all lose ($-8.9\%$ to $-1.5\%$; five significantly). The three homogeneous fleets, on which both schedulers are identical, differ by at most $1.3\%$. The losses occur under both condition orders and recur in the independent diversity sweep ($-6.7\%$ for two equal clusters, $t = -4.3$). These measurements do not identify the mechanism that distinguishes a single outlier from a paired or clustered minority. KV-cache usage was modest in these experiments, with eight sessions generating 128 tokens each, its peak size was at most 2.8\% of the model-weight size. We write ``$a@c$'' for $a$ sessions placed at cut layer $c$.

\begin{table}[t]
\centering
\fontsize{8}{9.5}\selectfont
\renewcommand{\arraystretch}{1.0}
\caption{DSB gain over exact-match batching by fleet composition
($N=8$, Llama-3.2-3B; six trials). Here, $a@c$ denotes $a$ sessions
at cut layer $c$, \emph{Tail} denotes $L-\max_i c_i$, and bold
indicates $|t|>2.5$.}
\label{tab:me-shape}

\begin{minipage}[t]{0.49\columnwidth}
\vspace{0pt}
\begin{tabular*}{\linewidth}
{@{\extracolsep{\fill}}lrr@{}}
\hline
\textbf{Fleet} & \textbf{Tail} & \textbf{Gain} \\
\hline

\multicolumn{3}{l}{\emph{Homogeneous}} \\
\hline
$8@2$  & 26 & $-1.3\%$ \\
$8@8$  & 20 & $+0.1\%$ \\
$8@14$ & 14 & $-0.7\%$ \\

\hline
\multicolumn{3}{l}{\emph{One outlier}} \\
\hline
$7@10,1@2$  & 18 & $\mathbf{+6.3\%}$ \\
$7@10,1@4$  & 18 & $\mathbf{+6.4\%}$ \\
$7@10,1@7$  & 18 & $\mathbf{+6.7\%}$ \\
$7@10,1@13$ & 15 & $+3.1\%$ \\
$7@10,1@16$ & 12 & $\mathbf{+16.6\%}$ \\
$7@10,1@19$ & 9  & $+5.8\%$ \\
$7@10,1@22$ & 6  & $+4.5\%$ \\
$7@10,1@26$ & 2  & $+2.2\%$ \\

\hline
\end{tabular*}
\end{minipage}
\hfill
\begin{minipage}[t]{0.49\columnwidth}
\vspace{0pt}
\begin{tabular*}{\linewidth}
{@{\extracolsep{\fill}}lrr@{}}
\hline
\textbf{Fleet} & \textbf{Tail} & \textbf{Gain} \\
\hline

\multicolumn{3}{l}{\emph{Clustered}} \\
\hline
$6@2,2@10$       & 18 & $\mathbf{-6.6\%}$ \\
$6@2,2@19$       & 9  & $\mathbf{-5.2\%}$ \\
$4@2,4@10$       & 18 & $-1.5\%$ \\
$4@2,4@18$       & 10 & $\mathbf{-5.6\%}$ \\
$6@2,1@10,1@22$  & 6  & $\mathbf{-3.6\%}$ \\
$3@2,3@12,2@22$  & 6  & $\mathbf{-8.9\%}$ \\

\hline
\multicolumn{3}{l}{\emph{Spread: one session per cut}} \\
\hline
$2,\ldots,9$
    & 19 & $\mathbf{+45.1\%}$ \\
\shortstack[l]{$2,4,7,10,$\\$13,16,19,22$}
    & 6 & $\mathbf{+17.6\%}$ \\

\hline
\end{tabular*}
\end{minipage}

\end{table}

\textbf{Operating envelope.} DSB is beneficial with at least four concurrent sessions, at least four distinct cut layers among them, a shared tail of approximately five or more layers, and contexts of up to about 2048 tokens. Outside this envelope, DSB can underperform exact-match batching, by up to $8.9\%$ on clustered fleets. The gain remains between $+78\%$ and $+84\%$ for contexts of up to 512 tokens, decreases to $+42.0\%$ at 2048 tokens, and is no longer significant at 4096 tokens (Table~\ref{tab:me-context}), because cache merging and splitting grow from $18\%$ to $43\%$ of each decode step as prompts lengthen from WildGPT-4 to LongReason. Relative to interleaving, the crossover depends on the arrival pattern (Table~\ref{tab:me-burst}). Under synchronized bursts on a fully spread fleet, interleaving leads with two and four sessions ($+18.2\%$ over DSB at four, $t = 20.7$), whereas DSB leads from six sessions ($+13.5\%$, $t = 22.3$) and by $+34.2\%$ at eight, where interleaving saturates near 122\,tok/s. Under continuous arrivals, DSB leads from $\bar{N}{=}4$, as shown above. The burst results also isolate the effect of the dispatcher: interleaving executes the same single-session calls as FIFO yet gains $+67\%$ to $+114\%$ over it, so this portion of the gain stems from a single scheduling thread rather than from batching.

\begin{table}[t]
\centering
\small
\caption{Decode throughput versus LongReason context length
(Llama-3.2-3B, $N=8$, cut layers 1--8, 20-layer shared tail; six trials).
Values are mean $\pm$ standard deviation; gain is relative to
exact-match batching.}
\label{tab:me-context}
\begin{tabular}{r|cc|c}
\hline
\textbf{Context} & \textbf{Exact-match} & \textbf{DSB} & \textbf{Gain} \\
\hline
64   & 86.0 $\pm$ 0.5 & 158.1 $\pm$ 4.4 & $+83.9\%$ \\
128  & 85.8 $\pm$ 0.5 & 154.1 $\pm$ 5.1 & $+79.7\%$ \\
256  & 84.7 $\pm$ 0.5 & 153.2 $\pm$ 3.8 & $+80.7\%$ \\
512  & 84.4 $\pm$ 0.5 & 150.5 $\pm$ 5.4 & $+78.2\%$ \\
1024 & 83.0 $\pm$ 0.5 & 137.4 $\pm$ 4.4 & $+65.5\%$ \\
2048 & 79.4 $\pm$ 0.6 & 112.7 $\pm$ 2.4 & $+42.0\%$ \\
4096 & 72.3 $\pm$ 0.4 & 75.4 $\pm$ 4.5  & $+4.2\%$ (n.s.) \\
\hline
\end{tabular}
\end{table}

\begin{table}[t]
\centering
\small
\setlength{\tabcolsep}{4pt}
\caption{Throughput under synchronized bursts with $N$ sessions at
distinct cut layers selected from 9 down to 2 (Llama-3.2-3B; 30 trials
with reversed scheduler order). Percentages are relative to FIFO; bold
denotes a significant best result ($|t|>2.5$).}
\label{tab:me-burst}
\begin{tabular}{c|cccc}
\hline
\textbf{N} & \textbf{FIFO} & \textbf{Exact-match} & \textbf{Interleave} & \textbf{DSB} \\
\hline
2 & 61.5 & 57.2 ($-7\%$)  & \textbf{102.8 (+67\%)}  & 52.7 ($-14\%$) \\
4 & 60.8 & 97.9 (+61\%)   & \textbf{121.6 (+100\%)} & 102.9 (+69\%) \\
6 & 58.8 & 107.6 (+83\%)  & 122.7 (+109\%)          & \textbf{139.3 (+137\%)} \\
8 & 57.0 & 112.4 (+97\%)  & 121.8 (+114\%)          & \textbf{163.5 (+187\%)} \\
\hline
\end{tabular}
\end{table}

\subsection{Discussion}

The two scales expose complementary constraints. Single-edge measurements show that nominal layer workload is only part of the latency. Idle gaps caused by edge computation and communication can slow the next GPU invocation. GPU dynamic voltage and frequency scaling (DVFS) and power-state transitions are plausible contributors, as prior work reports both inference sensitivity to DVFS and measurable frequency-switching costs~\cite{tang2019gpu_dvfs, han2025dvfs_aware_inference,velicka2025gpu_frequency_switching}. Clock locking may improve measurement consistency ~\cite{nvidia2026tensorrt_performance}, but our A6000 test did not eliminate the effect, and such controls also trade power for latency. Keeping the cloud supplied with useful work is therefore the more general system objective.

The multi-edge results address this objective while exposing a second form of heterogeneity. However, requests with different cut layers cannot use conventional exact-match batching. DSB recovers batching over their shared suffix and substantially improves throughput and latency under sufficient concurrency and cut-layer diversity. Its benefit is not universal, however; shallow shared tails, clustered fleets, or very long contexts can reduce or eliminate the gain. A practical scheduler should therefore consider both temporal readiness (to avoid idle-induced slowdown) and structural compatibility (to create a large shared tail), rather than selecting cut layers or batches from either factor alone.

The two testbeds isolate complementary effects. The single-edge study includes heterogeneous hardware and network links, whereas the multi-edge study uses a controlled server to attribute differences to scheduling and batching. Multi-edge prefill remains unbatched, and each session keeps its cut from prefill through decoding. These choices delimit the current measurements: they characterize scheduling with heterogeneous, fixed cuts, while wide-area multi-edge operation and online cut selection remain to be evaluated.

\section{Conclusion}

Layer-wise edge--cloud LLM inference allows devices with different capabilities to contribute to the same model, but also changes how a shared cloud GPU receives work. DySCo uses resident model shards to support configurable cuts without reloading weights, while keeping each session's KV caches with the layers that produce them. The single-edge measurements show that moving more layers to the edge does not necessarily reduce latency: the resulting idle interval can increase the lantency of the cloud's next forward pass. The multi-edge measurements reveal a related scheduling cost, as different cuts separate requests that share cloud suffix weights into distinct conventional batches. DSB recovers shared-suffix batching when concurrency and suffix depth permit, improving throughput and per-session latency in the evaluated workloads. The broader implication is that cut placement and cloud scheduling are coupled: a cut determines both when a session reaches the cloud and how much computation it can share with others.

Future work will evaluate geographically distributed edge fleets. It will also jointly choose cuts for new sessions and schedule their cloud work as device speed, network conditions, and request load change.

\bibliographystyle{IEEEtran}
\bibliography{bibliography}

\end{document}